# Accelerated identification of equilibrium structures of multicomponent inorganic crystals using machine learning potentials

Sungwoo Kang,[1,2] Wonseok Jeong,[1,2] Changho Hong,[1] Seungwoo Hwang,[1] Youngchae Yoon,[1] and Seungwu Han[1,*]

[1]Department of Materials Science and Engineering, Seoul National University, Seoul 08826, Korea.

[2]These authors contributed equally; Sungwoo Kang, Wonseok Jeong.

*e-mail: hansw@snu.ac.kr

## ABSTRACT

The discovery of new multicomponent inorganic compounds can provide direct solutions to many scientific and engineering challenges, yet the vast size of the uncharted material space dwarfs current synthesis throughput. While the computational crystal structure prediction is expected to mitigate this frustration, the NP-hardness and steep costs of density functional theory (DFT) calculations prohibit material exploration at scale. Herein, we introduce SPINNER, a highly efficient and reliable structure-prediction framework based on exhaustive random searches and evolutionary algorithms, which is completely free from empiricism. Empowered by accurate neural network potentials, the program can navigate the configuration space faster than DFT by more than $10^2$-fold. In blind tests on 60 ternary compositions diversely selected from the experimental database, SPINNER successfully identifies experimental (or theoretically more stable) phases for ~80% of materials within 5000 generations, entailing up to half a million structure evaluations for each composition. When benchmarked against previous data mining or DFT-based evolutionary predictions, SPINNER identifies more stable phases in the majority of cases. By developing a reliable and fast structure-prediction framework, this work opens the door to large-scale, unbounded computational exploration of undiscovered inorganic crystals.

Human history has evolved together with material innovation: steel production from heating iron with carbon triggered a shift from the Bronze to Iron age, and the growth of high-purity Si crystal ingots led to burgeoning of the computer age. In modern times, synthetic multicomponent materials are constantly developed to meet the demands of diverse applications. The Inorganic Crystal Structure Database (ICSD), including most of the experimentally synthesized inorganic compounds, has approximately 200,000 materials registered to date, and the data entry steadily increases by ~5,000 every year[1]. The sheer size and active expansion of the database reflect that new materials continue to drive scientific and engineering advances in fields such as electronics, energy harvesting/storage, and high-$T_c$ superconductors[2–7].

Despite the vast material library available today, it is far from being complete in ternary or higher-order (simply multinary hereafter) phases. Based on a rough estimate, only approximately 16% and 1% of ternary and quaternary compounds, respectively, are at least partially revealed[8]. This inspires reasonable hope that valuable materials can be discovered in the largely unexplored multinary domain. Furthermore, the current material repositories are chemically and synthetically biased. For instance, elemental occurrences within ternary compounds peak at oxygen with 22,476 counts in the ICSD, which is more than three times that of the next most frequent elements (Fe, Si, and S). This is mainly because oxygen is the most earth-abundant element and forms stable compounds with most metals. Thus, oxides are easier to synthesize than other compounds and benefit from synthetic recipes established throughout the long history. This indicates that the present material database is biased toward those with facile synthesis, possibly missing promising compounds that are unfamiliar today[9].

Considering the very large gap between current experimental throughput and the number of unknown materials, together with rising synthesis barriers, exploring the uncharted chemical space solely by experiment would be inefficient. Alternatively, experimental endeavors can take advantage of computational prescreening based on the density-functional theory (DFT) calculations. This has been demonstrated by a multitude of recent publications in which the discovery of new materials was accelerated by DFT: cathodes for Li-ion batteries[10], nitride semiconductors[11], metal nitrides[9], 18-electron compounds[12], and boron-based MAX phases[13]. Throughout these works, DFT results were able to suggest compositions that are presumably stable under ambient conditions and thus have high synthesizability. In addition, DFT predictions of basic properties could steer experimental resources to materials appropriate for specific applications.

Despite great promise, the current computational exploration of unknown materials inherently suffers from low throughput; in contrast to high-throughput screening of materials cataloged in the ICSD[14,15], the investigation into as-yet-unreported materials should start from the crystal structure prediction (CSP), preferentially for equilibrium phases with the lowest Gibbs free energy (evaluated at the DFT level) under given chemical composition and thermodynamic conditions. However, CSP is a classic NP-hard problem in which the computational load in identifying the global minimum exponentially increases with material complexity[16]. Compounded further by the high costs of DFT calculations, this results in extremely low throughput of DFT-based CSP.

Consequently, computational investigations are often limited to known prototypes[9,10] or short evolutionary steps[11,12], which risks resulting in metastable structures instead of the ground state. Reflecting this, the compositional territory has been scanned rather ‘conservatively’ over subsets close to known chemical families. If free energies can be evaluated much faster than with DFT, then more aggressive and far-reaching exploration will be viable on a large scale, raising the chance of discovering materials with novel functionalities.

Recently, machine-learning potentials have attracted considerable attention because they can provide DFT-level energies at a fraction of the cost, which finds immediate applications to CSP[17–20]. In applying machine-learning potentials to CSP, however, one is faced with difficulties in choosing training sets without information on crystal structures. We recently suggested a way to resolve this problem by using melt-quench molecular dynamics (MD) simulations[21]. The liquid simulation can self-start from a random distribution with rapid equilibration; thus, a priori information on the crystal structure is not required. In addition, the MD trajectory automatically samples diverse local orders that may appear in stable or metastable crystalline structures. We demonstrated that the Behler-Parrinello-type neural network potential (NNP)[22] trained with the melt-quench-annealing trajectories can serve as a high-fidelity surrogate model in CSP.

Going beyond the previous achievement, we herein develop a structure prediction framework combining NNPs with evolutionary or random searches, named SPINNER (Structure Prediction of Inorganic crystals using Neural Network potentials with Evolutionary and Random searches). Free from any empirical knowledge on material structures, the program identifies the global minimum in a brute-force style by harnessing the accuracy and speed of NNPs. In blind tests on ternary compounds with significant complexity and diverse crystal symmetries, SPINNER successfully identifies the experimental (or theoretically more stable) phases for ~80% of the cases.

The program also outperforms other favored approaches in the majority of test materials. The average computational throughput is ~4 days per one composition on a 36-core node, which is estimated to be $10^2$–$10^3$ times faster than pure DFT-based approaches. The outstanding performance of SPINNER allows for large-scale, unbounded computational exploration of undiscovered inorganic crystals.

The basic workflow of SPINNER is schematically presented in Fig. 1(a). In brief, for an input chemical composition (elements and stoichiometry), SPINNER first carries out a melt-quench-annealing simulation and trains an NNP with MD trajectories. To enhance the accuracy for ordered phases, the NNP is iteratively retrained over low-energy structures in the refining-state CSP (the upper part of Fig. 1(a)). In the main CSP proceeding up to 5000 generations (the lower part of Fig. 1(a)), SPINNER collects low-energy candidate structures within 50 meV atom$^{-1}$, which are finally sorted after full relaxations at the DFT level. For DFT calculations, we use the Vienna Ab initio Simulation Package (VASP)[23] with the Perdew-Burke-Ernzerhof (PBE) functional for exchange-correlation energies[24]. Figure 1b shows a schematic of the present evolutionary algorithms that are based on random generation, crossover, permutation, and lattice mutation. The random generation is heavily used over mutations, which we find to be instrumental in multi-component systems. Detailed descriptions of DFT calculations and evolutionary algorithms are provided in the Methods section. We stress that no empirical information (for instance, prototype data[25], bonding topology[26], or Pauling's rules[27]) is used in the present CSP scheme.

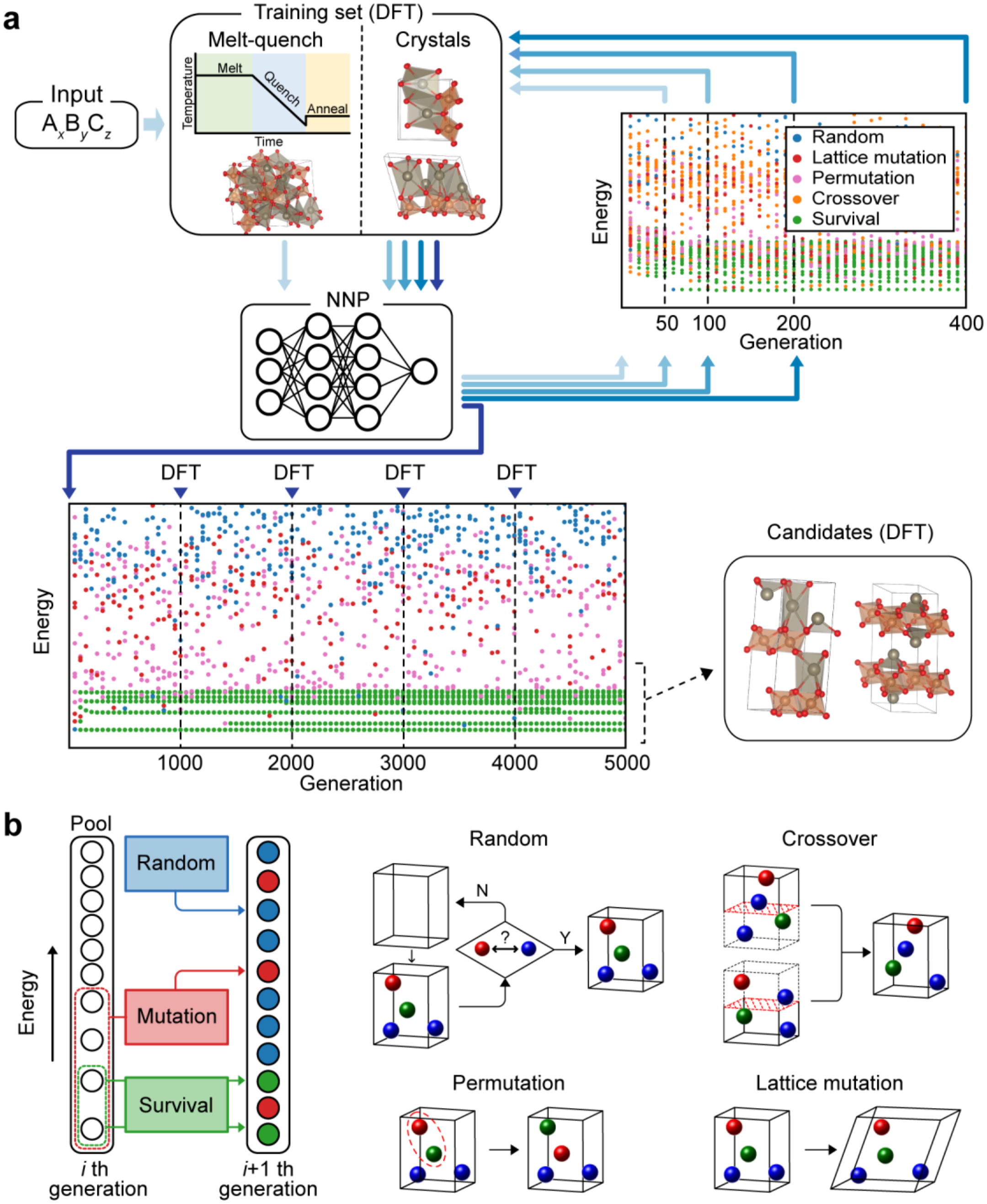

**Fig. 1. Schematic workflow of SPINNER.** (a) Schematic depiction of the whole workflow of CSP. Top left: An initial NNP is trained over disordered structures sampled from melt-quench-annealing trajectories of a given composition obtained by DFT calculations. Top right: The initial NNP is iteratively refined over low-energy crystal structures obtained during 400 generations of the structure search. Bottom left: With the refined NNP, the structure search is carried out up to 5000 generations. The quality of the NNP is monitored every 1000 generations. Bottom right: Free energies of final candidates are evaluated by DFT calculations. (b) Left: Representation of the evolutionary

algorithm. Under given fractions of random structure generation and mutations (crossover, permutation, and lattice mutation), new structures are generated for the next generation while low-energy structures additionally survive. Right: The schematic illustration of each generation method is presented. Decision chart in the random generation represents distance constraints.

## Results

**Selection of test materials.** In benchmarking search algorithms tackling NP-hard problems, the global minimum is usually unknown. The situation is slightly different in CSP because the equilibrium structures that are experimentally resolved by diffraction analysis mostly equate to the global minimum for the given chemical formula and thermodynamic conditions. Therefore, blind tests with the compositions reported in the ICSD can serve as an ultimate evaluation of the performance of a CSP algorithm. Here we focus on ternary compounds because the corresponding database spans a wide range of chemistries and structural prototypes.

Within the ICSD, we first select ternary compounds measured at room temperature or below and at atmospheric pressure, as we are mainly concerned with materials that are stable under ambient conditions. We also consider only high-quality ($R$ < 0.1) ordered crystals with well-defined structures and rule out molecular crystals as well as compounds including 3d transition elements (V–Zn), lanthanides and actinides. In the latter materials with 3d and f electrons, the magnetic ordering influences the free energy, but the current implementation of NNPs has yet to resolve the fine energy scales of magnetic interactions. (We note that meaningful developments are underway[28,29].) When distinct crystal structures with the same composition are available in the ICSD, the most stable structure within the PBE functional is regarded as the reference. Among the filtered structures, we randomly select 50 materials under the condition that at least one crystal is selected from the 32 crystallographic point groups (omitting $6/m$ due to zero occurrence), which secures diversity of the test structures. To exclude structures that are too simple, we choose compounds with the formula units ($Z$) in the primitive cell being at least 4. (So, the numbers of atoms are at least 12). We additionally handpick 10 structures to further diversify local motifs and chemistry. The full list of the selected 60 materials is provided in the Supplementary Information. Regarding the band gap ($E_g$), there are 18 metals, 13 semiconductors ($0 < E_g \leq 2$ eV), and 29 insulators ($E_g > 2$ eV). (The finite band gaps are calculated within one-shot hybrid functional calculations[30,31].) Most structures have a hull energy of 0 in the Materials Project database[32] (see Supplementary Table 1).

**Structure prediction by SPINNER.** In running SPINNER, we assume the same $Z$ value as in the ICSD structure. Fig. 2(a) shows $\Delta E_{min}$, the offset of the PBE energy of the most stable structure after 5000 generations of structure search from that of the experimental structure. Since the pool size is 24–80 (see the Methods section), up to half a million structure relaxations are performed for each composition. The color code of the data indicates the earliest generation at which the minimum energy structure is identified ($N_g$). For 75% of the compositions (45 of 60), SPINNER predicts the reference ($\Delta E_{min} = 0$) or a lower-energy ($\Delta E_{min} < 0$) structure, which is mostly found within 1000 generations (38 of 45). The largest error occurs for $Sr_2Pt_3In_4$ with a $\Delta E_{min}$ of 36 meV atom$^{-1}$. Figures at the bottom of Fig. 2(a) display the unit cells of some materials with $\Delta E_{min} = 0$.

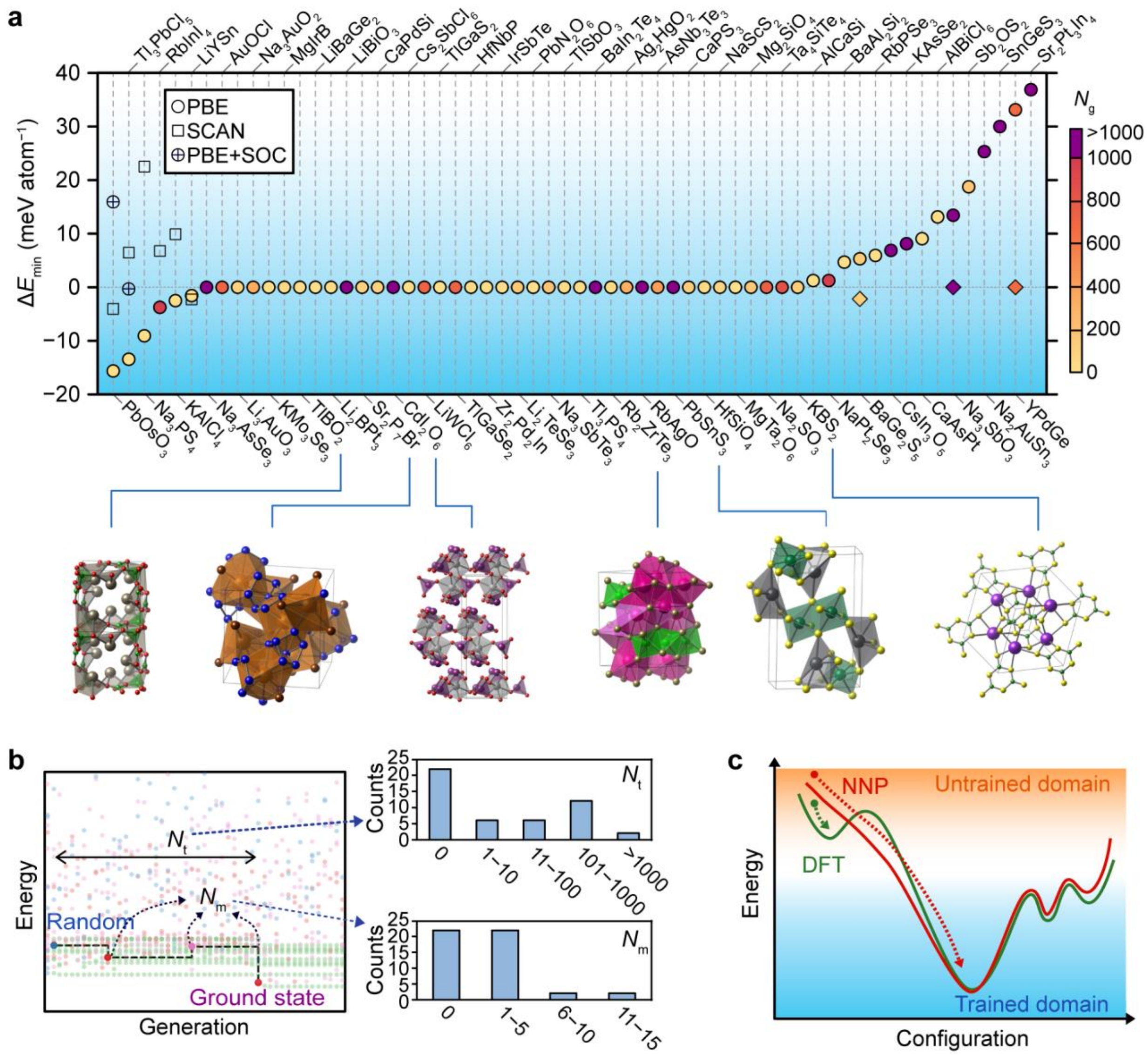

**Fig. 2. Search results of SPINNER against structures in the ICSD.** (a) Distribution of the energy difference between the most stable structure found by SPINNER and the ICSD structure ($\Delta E_{\mathrm{min}}$). The color codes indicate the generation at which the lowest-energy structure was identified ($N_{\mathrm{g}}$). The circle, square, and sun cross indicate the exchange-correlation functional and diamonds are results obtained with the conventional cell. Equilibrium structures of some compositions with $\Delta E_{\mathrm{min}} = 0$ are displayed below. (b) Definition and distribution of $N_{\mathrm{t}}$ and $N_{\mathrm{m}}$ for materials with $\Delta E_{\mathrm{min}} \leq 0$; $N_{\mathrm{t}}$ denotes the number of generations that elapsed from random seeding until the ground state is identified, and $N_{\mathrm{m}}$ is the number of mutations involved in this process. $N_{\mathrm{t}} = 0$ means that the ground state is directly obtained by relaxation from a random structure. (c) Schematic illustration of potential energy surfaces of the DFT and NNP.

In Fig. 2(a), six materials have negative $\Delta E_{\min}$ down to −16 meV atom$^{-1}$, meaning that the structure found by SPINNER is more stable than the experimental phase within the PBE functional. We note that both structures share almost the same local order, differing in ranges beyond ~4 Å. Since the ICSD does not register any other experimental structures for these compositions, the energy ordering obtained by the PBE functional is likely incorrect, calling for further investigation. In ref. [33], the PBE functional was found to incorrectly stabilize metastable phases for some binary materials, which was partly addressed by the SCAN functional[34,35]. The empty squares in Fig. 2(a) are the $\Delta E_{\min}$'s recalculated by the SCAN functional. Except for $PbOsO_3$ and LiYSn, the $\Delta E_{\min}$ values becomes positive. This confirms that the SCAN functional is more accurate than PBE in ordering low-energy phases. Since heavy elements such as Tl, Pb, and Bi possess large spin-orbit coupling (SOC), we also recalculate energies for $Tl_3PbCl_5$ and $PbOsO_3$ with SOC (see sun crosses) and find that the $\Delta E_{\min}$ for $Tl_3PbCl_5$ and $PbOsO_3$ increases to −0.2 and 16 meV atom$^{-1}$, respectively. Thus, the incorrect energy orderings are mostly rectified by introducing more sophisticated energy functionals. (We note that the PBE functional without SOC still correctly reproduce the equilibrium phases in many compositions including Tl, Pb, and Bi.) Except for $Na_3PS_4$ and $KAlCl_4$, experimental structures are not found among the final candidate structures for materials with $\Delta E_{\min}$ < 0. Even though the energies of the ICSD structures are close to the global minimum, they compete with other metastable structures populating at similar energies, lowering the chance of appearing during given generations. This indicates the importance of the appropriate exchange-correlation functional in preparing training data to efficiently identify the ground state. However, the SCAN functional or SOC significantly increases the computational time for the DFT part.

We assess the inference accuracy of NNP based on two metrics: The first is the absolute energy difference ($\Delta E_0$) between DFT and NNP for the experimental structure (relaxed within each method), which relates to how well NNP predicts the structure and energy of the reference structure. The second metric ($\Delta\bar{E}$) is similar to the first one but it is averaged over final candidates within bottom 50 meV atom$^{-1}$. $\Delta\bar{E}$ indicates how accurately NNP ranks the energies of stable and metastable phases. $\Delta E_0$ and $\Delta\bar{E}$ are provided for every material in Supplementary Table 1. When averaged over materials with $\Delta E_{\mathrm{min}} \leq 0$, $\Delta E_0$ and $\Delta\bar{E}$ are 12.9 and 11.8 meV atom$^{-1}$, respectively, meaning the potential energy surfaces of DFT and NNP agree well around the equilibrium and low-energy metastable structures. This confirms that the trained NNPs are good surrogate models of DFT. The remaining errors are partly attributed to a low resolution in describing medium- to long-order beyond cutoff radii of descriptors, which are 6 and 4.5 Å for radial and angular parts, respectively. This can be improved by increasing the cutoff radius, but it adversely affects the computational efficiency due to rising costs in evaluating descriptors. The analysis on the failed cases ($\Delta E_{\mathrm{min}} > 0$) is given in the Discussion section.

To understand how SPINNER effectively addresses the NP-hardness, we analyze the evolutionary steps leading to the experimental structure for materials with $\Delta E_{\mathrm{min}} \leq 0$ based on two parameters, $N_{\mathrm{t}}$ and $N_{\mathrm{m}}$. As schematically shown in Fig. 2(b), $N_{\mathrm{t}}$ counts the generations from random seeding to the appearance of the equilibrium structure at $N_{\mathrm{g}}$. $N_{\mathrm{m}}$ indicates the number of mutations within $N_{\mathrm{t}}$ steps. The distribution of $N_{\mathrm{t}}$ in Fig. 2(b) shows that $N_{\mathrm{t}}$ is 0 for almost half of the cases, meaning that the ground-state structure is obtained directly from relaxing a random structure, without any mutations involved. Even in cases undergoing finite mutations, $N_{\mathrm{m}}$ is mostly within 5. Intriguingly, the randomly generated structures are very close to the global minimum. There are two reasons contributing to this. First, pair-wise minimum distances obtained from melt-quench-annealing trajectories are imposed as constraints in the structure generation and relaxation (see the Methods section). This condition filters out ~99% of the randomly generated structures that are unlikely to relax into low-energy structures (see the decision tree in Fig. 1(b)). It also effectively protects atomic configurations from evolving into unphysical structures during relaxation. Note that the minimum distance constraints often extend to 2–3 Å, which are much stronger than simple conditions preventing too short bonds. Second, we find that when the random structures leading to the global minimum are relaxed by DFT instead of the NNP, more than half of them relax to high-energy metastable structures instead of the global minimum. This implies that the energy landscape around the global minimum is different between DFT and the NNP (see Fig. 2(c)). That is, DFT adaptively forms chemical bonds that can locally stabilize the starting configuration; thus, the search can be stuck at a high-energy metastable structure. Having not machine-learned such chemical bonds, the NNP unwittingly smooths out the corresponding energy region, directing the structure to the global minimum. Figuratively, the NNP "catalyzes" the relaxation process to the equilibrium, eliminating or lowering energy barriers around certain metastable structures. This also effectively increases the volume of the configurations that can funnel down to the equilibrium structure.

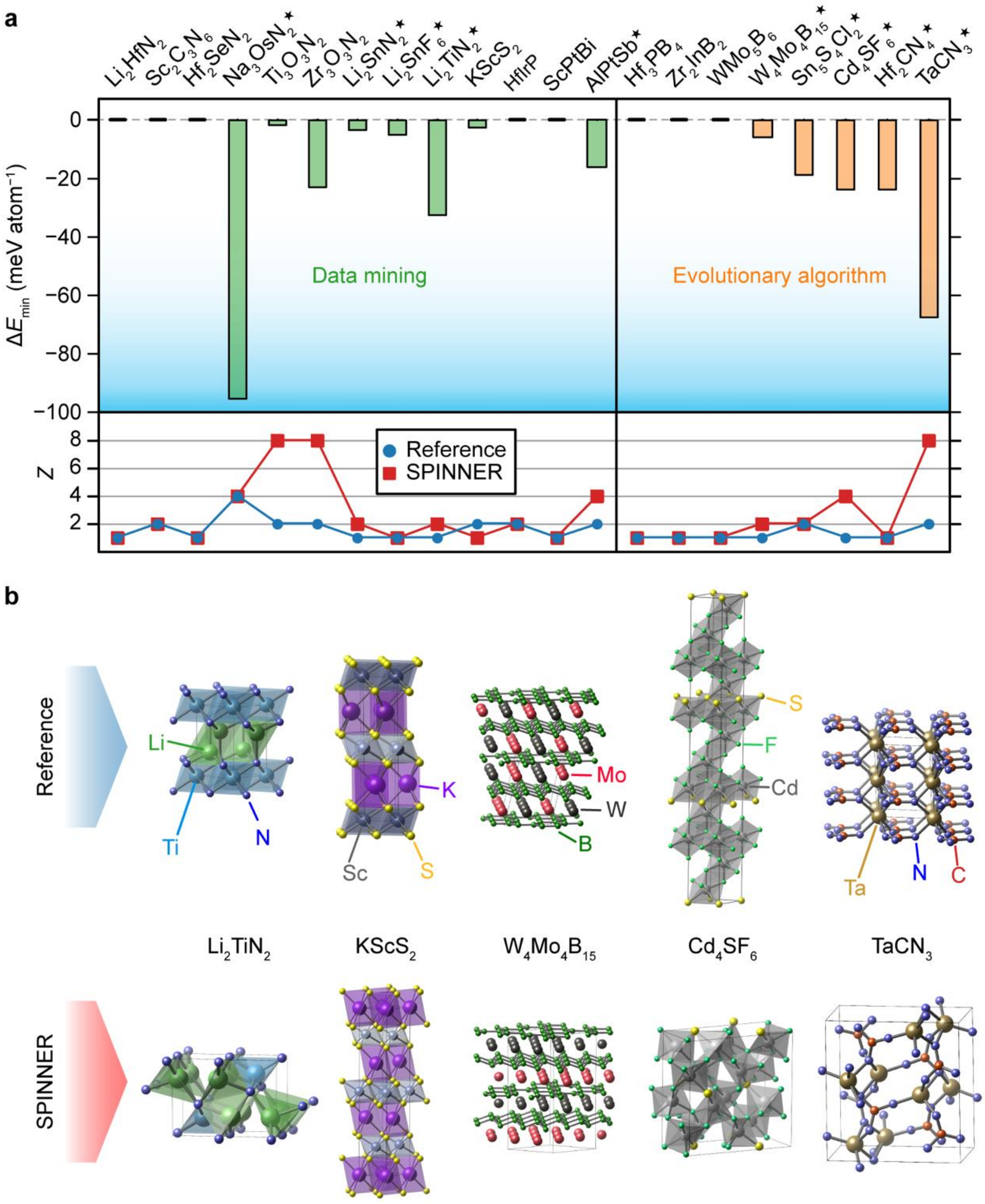

**Fig. 3. Benchmark results of SPINNER against other methods.** (a) Upper part shows energy difference between the structure predicted by SPINNER and the reference structure ($\Delta E_{\mathrm{min}}$) predicted by data mining or evolutionary algorithms. The lower part compares the formula unit in the unit cell ($Z$) between the reference and the value at which the lowest energy is found by SPINNER. The structures are adopted from ref. [9] ($Li_2HfN_2$, $Sc_2C_3N_6$, $Hf_2SeN_2$, $Na_3OsN_2$), ref. [36] ($Ti_3O_3N_2$, $Zr_3O_3N_2$), ref. [37] ($Li_2SnN_2$, $Li_2SnF_6$, $Li_2TiN_2$), ref. [38] ($KScS_2$), ref. [12] (HfIrP, ScPtBi, AlPtSb), ref. [39]

($Hf_3PB_4$, $Zr_2InB_2$), ref. [40] ($WMo_5B_6$, $W_4Mo_4B_{15}$), ref. [41] ($Sn_5S_4Cl_2$, $Cd_4SF_6$), and ref. [42] ($Hf_2CN_4$, $TaCN_3$). The chemical formulas with star marks do not have corresponding structural prototypes in the ICSD. (b) Examples illustrating the structural difference between the crystal structures in references (top) and ones identified by SPINNER (bottom).

**Comparison with other approaches.** To benchmark SPINNER against other CSP methods, we select ternary structures from the literature that were theoretically predicted by either data-mining known prototypes[9,12,36–38] or using DFT-based evolutionary approaches such as a genetic algorithm[39–41] or particle swarm optimization[42]. The materials are listed in Fig. 3(a), and understandably, they are all non-oxides (nitrides, borides, etc). None of these compositions has ever been synthesized as far as we are aware except for $KScS_2$, $Sc_2C_3N_6$, and ScPtBi (see below). For these materials, we perform CSP with SPINNER up to 1000 generations with $Z$ ranging from 2 to 8. The plot at the top of Fig. 3(a) shows $\Delta E_{\mathrm{min}}$ in reference to the structure in the literature. (The structures from the references are fully relaxed within the present DFT method.) SPINNER identifies lower-energy structures for the majority of cases (13 of 21) and the same structures for the rest. In ref. [41], CSP was performed with the PBEsol functional[43] instead of PBE for $Sn_5S_4Cl_2$ and $Cd_4SF_6$. As a comparison, we evaluate $\Delta E_{\mathrm{min}}$ with PBEsol and find that compounds identified in this study are still more stable than those predicted in the reference by 31 and 4 meV atom$^{-1}$ for $Sn_5S_4Cl_2$ and $Cd_4SF_6$, respectively. The lower part of Fig. 3(a) shows $Z$ values of the primitive cells with the lowest energies (solid squares) in comparison with those in the references (solid circles). Lower energies are found mostly at $Z$ values higher than those in the literature. In addition, SPINNER finds the lowest-energy structure often at a larger $Z$ than that of the identified primitive cell. This indicates the importance of trials with diverse $Z$ numbers. The NNP is far more advantageous in varying $Z$ than DFT owing to the linear scaling with respect to the number of atoms, in contrast to the cubic scaling of DFT.

While structures found by SPINNER have usually local structures similar to reference structures (such as $KScS_2$ and $W_4Mo_4B_{15}$ in Fig. 3(b)), different short-range orders appear in materials such as $Li_2TiN_2$, $Cd_4SF_6$ and $TaCN_3$ (see Fig. 3(b)). We note that $Cd_4SF_6$ and $TaCN_3$ were discovered by DFT-based evolutionary searches[41,42]. The failure of the previous works in identifying the present lower-energy structures might be attributed to small $Z$ numbers and/or few generations. In Fig. 3(a), $Na_3OsN_2$ shows the biggest energy drop. Both structures are similar in local order, but the reference structure is distorted from that found by SPINNER. We comment that metastable structures can have distinct physical properties compared to the ground state. In the example of $HfO_2$, the high-temperature tetragonal phase that is metastable by 57 meV atom$^{-1}$ at 0 K possesses a far higher dielectric constant (70) than that for the low-temperature monoclinic phase (16)[44]. This underscores the importance of identifying the true global minimum for accurate prediction of material properties.

For the 13 structures newly identified in Fig. 3(a) by SPINNER, we check the existence of corresponding prototypes in the ICSD using AFLOW-XtalFinder[45]. We find that 10 of 13 materials do not have any matching prototypes (see star marks in Fig. 3(a)). This suggests that the current prototypes for the ternary phase are not sufficiently cataloged, which is understandable because materials such as nitrides have been synthesized much less frequently than oxides. Notably, the prototype of the SPINNER-identified $Zr_3O_3N_2$ and $Ti_3O_3N_2$ is commonly $Ti_3O_5$. We think that ref. [36] missed this prototype because they did not fully consider partial ion-exchanges replacing O with N. In passing, the three compositions for which SPINNER and previous evolutionary searches agree have corresponding prototypes in the ICSD.

$Sc_2C_3N_6$ was recently synthesized with the structure predicted by data mining[46], which was confirmed by SPINNER. For $KScS_2$, there was an experimental report[47] preceding the theoretical prediction, but it was not recognized by ref. [38]. The experimental crystal structure is consistent with the present work, which is more stable than that for ref. [38] by 2.5 meV $atom^{-1}$(see Fig. 3(b)). The crystal structure of ScPtBi was predicted and confirmed by synthesis in ref. [12]. Although it was synthesized as a multiphase, SPINNER does not identify any phases within 50 meV $atom^{-1}$ other than the one predicted in ref. [12]. (Possibly, SOC would be necessary due to Pt and Bi.)

## Discussion

Regarding the computational cost, the whole computation from the melt-quench-annealing MD to 5000 CSP steps takes 3–5 days on a 36-core Amazon® server (CPU of Intel® Xeon Platinum 8000-series). (For 1000 evolution steps, it takes 2–3 days.) This excludes human time in managing data flow and making decisions, which can be executed instantly by automation (under development). On average, the workload can be split into the DFT-MD (25%), training of the NNP (10%), SPINNER (60%), and DFT relaxation of crystals (5%). The computational time for the DFT part varies widely depending on the number of electrons. The SPINNER part is highly scalable on massively-parallel computers. As a specific example of $LiWCl_6$ tested above, it takes 84 hours for 5000 generations with the pool size of 64 and $Z$ = 4. When tested with the same computational resources and conditions, a DFT evolutionary program[48] could proceed up to only 6 generations under the setting suggested in the manual. Therefore, the entire 5000 generations would take several years with DFT. (We add that actual generations required for identifying the equilibrium phase can be different between DFT and NNP.) In terms of monetary price, the structure prediction of one material by SPINNER costs approximately 200 US dollars based on the current pricing policy of the Amazon® server. This estimation indicates that it would be viable to construct databases of ~1000 materials targeted for a specific application at a reasonable cost and in a reasonable time frame.

Despite impressive performance, SPINNER failed with 25% of test ternary materials, which merits further discussion. First, $\Delta E_0$ and $\Delta \bar{E}$ averaged over materials with $\Delta E_{\mathrm{min}} > 0$ (see Supplementary Table 1) are 41.0 and 43.3 meV atom$^{-1}$, which are substantially larger than 12.5 and 11.9 meV atom$^{-1}$ for structures with $\Delta E_{\mathrm{min}} \leq 0$. Most notably, the NNPs for $SnGeS_3$ and $Sr_2Pt_3In_4$ show the largest $\Delta E_0$ ($\Delta \bar{E}$) of 75.5 (227.3) and 83.5 (140.0) meV atom$^{-1}$, respectively. For these materials, locating the equilibrium structures would be very difficult, as they are high in the energy scale. The origin of the poor NNP quality requires further investigation. (Increasing the cutoff radii of descriptors was not helpful.) Second, among the failed cases, qualities of the NNP are acceptable in several materials, with accuracy metrics on par with those with $\Delta E_{\mathrm{min}} \leq 0$. In these materials, the experimental structures would be found in principle by extending the evolutionary process beyond 5000 generations. We note that $BaGe_2S_5$, $Na_3SbO_3$, and YPdGe have rather flat or pointed cell shapes. Such low-dimension-like geometries are entropically low with small chances to occur in the random generation. Since conventional unit cells for these materials have more 3-D-like structures than the primitive cells, we additionally carry out a structure search with $Z$ equal to that of the conventional cells (see Supplementary Table 1). In all three cases, SPINNER successfully identifies experimental or theoretically more stable structures (see diamonds in Fig. 2(a)). This again highlights the importance of multiple trials at different $Z$ numbers.

Even though the training set is constructed within a specific stoichiometry, dynamic fluctuations during the melt-quench-annealing process extend transferability of NNPs. For example, employing the NNP developed for $Mg_2SiO_4$, we try CSP for $MgSiO_3$ and SPINNER successfully identifies the experimental crystal structure (ICSD-ID of 196432). (Note that both materials have the same valence states for every element.) Since the NNP training set has a single stoichiometry, atomic-energy offsets among chemical species become arbitrary, resulting in constant energy shifts between the DFT and NNP energies for crystalline structures of $MgSiO_3$[49]. Nevertheless, energy ordering among the low-energy structures is well maintained, leading to the successful identification of the equilibrium phase. Such extended transferability would be advantageous in saving training sets for a variable-composition crystal search[40], which would require multiple MD simulations covering a range of compositions.

To test SPINNER on compounds other than ternary phases, we carry out similar studies on B, $TiO_2$, $P_3N_5$, $NbPd_3$, $Li_{10}GeP_2S_{12}$[2], and $InGaZnO_4$[3] whose equilibrium structures are experimentally known. In all cases, ground-state structures are found within or around 5000 generations. In detail, B is well known for a complicated bonding nature and rich polymorphism[17,50]. The ground-state structure ($\alpha$-$B_{12}$) is found at 5055 generations for $Z$ = 12 (primitive cell), while it is found at 748 steps for $Z$ = 36 (conventional cell). Similar to a previous study[50], the training error is relatively large (~30 meV atom$^{-1}$) but the energy ordering among low-energy structures is good enough to predict the equilibrium phase correctly. For $TiO_2$, we try $Z$ = 8 and obtain all the major polymorphs such as rutile, anatase, and brookite among the final candidate structures. Even though the test sets are small compared to the ternary compounds, these results support that the present CSP framework works regardless of material complexity. However, quaternary or higher-order compounds may generally require longer generations than in the present work, which needs further investigation.

## Methods

**Training machine-learning potentials.** The NNPs used in CSP are first trained with DFT-MD simulations on the melt-quench-annealing trajectory of each compound. All DFT calculations are performed with the Vienna Ab initio Simulation Package (VASP)[23] and the Perdew-Burke-Ernzerhof (PBE) functional is used for the exchange-correlation energy of electrons[24]. The initial structures are prepared by randomly distributing ~80 atoms under the given composition and superheating them at 4000 K for 4.5 ps. Next, we obtain liquid-phase trajectories for 16 ps at ad hoc melting temperature ($T_m$). Subsequently, the liquid is quenched at a cooling rate of 200 K ps$^{-1}$ from $T_m$ to 300 K and then annealed at 500 K for 4 ps to sample amorphous structures. The cutoff energies, **k**-point meshes, cell volumes, and $T_m$ are determined following the scheme explained in ref. [21]. As an alternative to the melt-quench-annealing simulation, we find that the metadynamics in the atomic environment space[51] can also generate trajectories that can be used as training sets for CSP. (However, computational costs are higher.)

For training NNPs, we use the SIMPLE-NN package[52]. Behler-Parrinello-type symmetry function vectors are adopted as input features. For each pair of atomic species, 8 radial and 18 angular components are used with cutoff radii of 6 and 4.5 Å, respectively[53]. We train the NNPs until the root mean square errors of the validation set reduce to within 20 meV atom$^{-1}$, 0.3 eV/Å, and 20 kbar for the energy, force, and stress components, respectively. Other parameters for NNPs are identical to ref. [21].

To enhance accuracy on ordered phases, the NNPs are iteratively retrained over low-energy structures in the refining-state CSP (the upper part of Fig. 1(a)). With the NNPs trained by the melt-quench-annealing trajectories, we carry out CSP for 50 generations and select 10 structures: 5 lowest-energy structures and 5 structures with the lowest antiseed weights (see below) within bottom 100 meV atom$^{-1}$. These structures are further relaxed within DFT by using AMP$^2$ [30]. The relaxation terminates when atomic forces and stress tensors are less than 0.1 eV/Å and 20 kbar, respectively. The atomic positions and energies are sampled during the DFT relaxation, which are augmented to the training set for refining the NNPs. We find that the initial cell structure sometimes undergoes significant deformations, which necessitates adapting the **k**-point mesh consistently. To this end, the **k**-point mesh is checked and modified every 10 relaxation steps according to the converged **k**-point spacing. Then, we continue CSP and retrain the NNPs at 100, 200, and 400 generations in similar ways but excluding structures already learned in the previous iterations. In total, 40 crystal structures are sampled to refine the NNPs.

**Workflow of crystal structure prediction.** With the refined NNPs, we perform up to 5000 generations of the main CSP (the lower part of Fig. 1(a)). The structures are generated by random seeding, permutation, lattice mutation, and crossover algorithms. Detailed descriptions of each mode are provided in the next subsection. The generated structures are relaxed within the NNPs using the LAMMPS code[54]. We use the restraint option during structure relaxation, which prevents atom pairs from being closer than a certain distance limit through repulsive harmonic forces. The pair-wise distance limits are set to the minimum values found for the corresponding pair during the melt-quench-annealing process. The restraint option effectively prohibits the structures from evolving into untrained domain. In comparison, the structural relaxation by DFT methods typically takes $10^2$–$10^3$ times that of the NNP.

To prevent unphysical structures from appearing as candidate structures, we compare the single-shot DFT energy and NNP energy of the lowest-energy structure every 1000 generations. If the discrepancy is greater than 50 meV atom$^{-1}$, then the NNP is retrained over 10 structures selected similarly to the refining stage (see above). Last, we perform DFT structure relaxation for final candidates that lie within bottom 50 meV atom$^{-1}$ using the AMP[2]. For accurate evaluation of energies, we adopt tight convergence criteria of 2 meV atom$^{-1}$ and 3 kbar for energy and pressure, respectively.

**Evolutionary algorithms.** Following ref. [48], SPINNER utilizes evolutionary algorithms to find the global minimum in the configuration-energy space under the given composition and *Z* number. From extensive tests, we find that random generation, crossover, permutation, and lattice mutation algorithms are particularly effective in the structure prediction of multinary phases, and so the current version of SPINNER is based on these four generation modes. In generating random structures, the space group, lattice vectors, and Wyckoff positions are selected randomly using the RandSpg package[55]. The atomic density in the first generation is set to that of the amorphous phase generated by the DFT-MD simulations. Then, the volume of each structure in the *i*th generation ($i > 1$) is chosen randomly between 70% and 130% of the volume of the lowest-energy structure in the previous generation. Distance constraints are imposed on all pairs of atoms such that distances between pairs are longer than the minimum values found for the corresponding pair during the melt-quench-annealing process. In applying the crossover algorithm, we note that the conventional approach often unnecessarily disrupts stable chemical units (for instance $SiO_4$ tetrahedra in $Mg_2SiO_4$). To avoid this, SPINNER utilizes atomic energies of the NNP in choosing cutting planes such that the average atomic energy within the slab is sufficiently low. The average atomic energies are also used in adjusting lattice vectors and atomic registries in combining the two slabs. This effectively conserves chemically-stable local units.

During the evolutionary steps, the structural similarity is measured by the partial radial distribution function (PRDF)[56] and if two structures are determined to be too similar, then one of them is discarded from the pool. We also adopt an antiseed algorithm to diversify the structure pool[57]. The antiseed weight of $i$th structure ($A_i$) is defined as follows:

$$A_i = \sum_a \exp(-\frac{{d_{ia}}^2}{2\sigma^2}), \quad (1)$$

where the summation runs over structures in the inheritable pool (bottom 200 meV atom$^{-1}$ in the refining stage and 100 meV atom$^{-1}$ during the main CSP), $d_{ia}$ is the distance between the $i$th and $a$th structures measured by PRDF, and $\sigma$ is a Gaussian width. Unlike the original scheme[57], we use the antiseed weight in defining the probability ($P_i \propto \exp(-A_i/\sigma_A)$ where $\sigma_A$ is a parameter) for a structure to be inherited in the next generation by mutation (see below).

The number of structures generated by the operations is set to twice the number of atoms in the simulation cell. The inherited structures are chosen within bottom 200 and 100 meV atom$^{-1}$ for the refining-stage and main CSP, respectively, and we choose half of them randomly and the rest are chosen according to the probability distribution of $\{P_i\}$ defined in the above. In the refining stage, ratios of operations of random generation, crossover, permutations, and lattice mutations are 30%, 50%, 10%, and 10%, respectively. The crossover tends to generate diverse local orders, which is useful in training robust NNPs. However, the algorithm is less effective in finding the ground state due to the low symmetries of the generated structures, so we exclude the crossover operation in the main CSP in which random (70%), permutation (20%), and lattice mutation (10%) are employed. In addition to the structural pool generated by these operations, we carry low-energy structures in the bottom 100 and 50 meV atom$^{-1}$ to the next generation in the refining stage and main CSP, respectively.

## Data availability

Except for materials already available in the literature, ICSD, and the Materials Project, all of the lowest-energy structures found by SPINNER are uploaded to the Figshare Repository[58]. They are also available at SNUMAT[59] together with basic DFT results such as band gaps. SNUMAT supports RESTful API for search and download.

## Code availability

The core part of SPINNER is available at https://github.com/MDIL-SNU/SPINNER. This program carries out the evolutionary structure search based on the given NNP. SIMPLE-NN for training NNPs is open at https://github.com/MDIL-SNU/SIMPLE-NN. A fully automated end-to-end version of SPINNER is under development.

**Acknowledgements**

This work was supported by Samsung Electronics Co., LTD and Korea Institute of Ceramic Engineering and Technology (KICET) (N0002599). The computations were carried out at the Korea Institute of Science and Technology Information (KISTI) supercomputing center (KSC-2020-CRE-0125). We thank Youngho Kang, Kanghoon Yim, and Yong Youn for critical comments.

# Supplementary Information for "Accelerated identification of equilibrium structures of ternary compounds using machine learning potentials"

Sungwoo Kang,[1,2] Wonseok Jeong,[1,2] Changho Hong,[1] Seungwoo Hwang,[1] Youngchae Yoon,[1] and Seungwu Han[1,*]

[1]Department of Materials Science and Engineering, Seoul National University, Seoul 08826, Korea

[2]These authors contributed equally; Sungwoo Kang, Wonseok Jeong

**Corresponding Author**

*E-mail: hansw@snu.ac.kr

**Supplementary Table 1.** The table provides information and results on test compounds in Fig. 2a. The columns under ICSD are data on the equilibrium phase in the ICSD. $Z$ and $N_{at}$ are the numbers of formula units and atoms in the unit cell, respectively. Band gaps ($E_g$'s) are calculated by one-shot hybrid functional calculations[1] and hull energies ($E_{hull}$'s) are cited from the Materials Project.[2]

For the symbols under SPINNER, we refer to the main text. Under $\Delta E_{min}$ we write the energy difference between the most stable structure found by SPINNER within ICSD structure primitive cell size and the ICSD structure calculated by PBE functional. When needed, we additionally write the energy difference calculated by SCAN functional (marked by †), or by PBE when the spin-orbit coupling is considered (‡). Also, the numbers with the * mark indicate the energy difference between the most stable structure found within ICSD structure conventional cell size and the ICSD structure. The unit for $E_{hull}$, $N_g$, $\Delta E_0$, $\Delta\bar{E}$, and $\Delta E_{min}$ are meV atom$^{-1}$.

| Formula | ICSD | | | | | | SPINNER | | | |
|---|---|---|---|---|---|---|---|---|---|---|
| | ID | Point group | $Z$ | $N_{at}$ | $E_g$ (eV) | $E_{hull}$ | $N_g$ | $\Delta E_0$ | $\Delta\bar{E}$ | $\Delta E_{min}$ |
| $PbOsO_3$ | 23444 | $m\bar{3}m$ | 4 | 20 | 0 | 0 | 5 | 64.1 | 20.3 | –15.6/ –4.0$^{\dagger}$/16.0$^{\ddagger}$ |
| $Tl_3PbCl_5$ | 1262 | 4 | 4 | 36 | 6.0 | 9 | 11 | 3.8 | 5.9 | –13.6/ 6.0$^{\dagger}$/ -0.2$^{\ddagger}$ |
| $Na_3PS_4$ | 72860 | $\bar{4}2m$ | 2 | 16 | 3.3 | 0 | 1 | 8.1 | 16.4 | –9.1/22.3$^{\dagger}$ |
| $RbInI_4$ | 36601 | 3m | 6 | 36 | 3.5 | 0 | 990 | 2.9 | 11.9 | –4.0/6.4$^{\dagger}$ |
| $KAlCl_4$ | 1704 | 2 | 4 | 24 | 6.8 | 0 | 238 | 2.9 | 10.1 | –2.7/9.5$^{\dagger}$ |
| LiYSn | 32041 | 6mm | 8 | 24 | 0 | 0 | 13 | 3.3 | 12.6 | –1.7/–2.0$^{\dagger}$ |
| $Na_3AsSe_3$ | 50491 | 23 | 4 | 28 | 2.8 | 0 | 4004 | 1.4 | 7.3 | 0 |
| AuOCl | 8190 | $\bar{3}$ | 6 | 18 | 2.3 | 0 | 723 | 22.0 | 20.1 | 0 |
| $Li_3AuO_3$ | 15113 | 4/mmm | 4 | 28 | 3.5 | 0 | 80 | 9.4 | 11.2 | 0 |
| $Na_3AuO_2$ | 62066 | 4/mmm | 4 | 24 | 3.1 | 0 | 344 | 2.5 | 3.1 | 0 |
| $KMo_3Se_3$ | 603628 | 6/m | 2 | 14 | 0 | 0 | 6 | 2.1 | 4.8 | 0 |
| MgIrB | 409979 | 622 | 6 | 18 | 0 | 0 | 46 | 1.1 | 4.3 | 0 |
| $TlBO_2$ | 36404 | 4 | 8 | 32 | 3.8 | 0 | 5 | 0.1 | 4.7 | 0 |
| $LiBaGe_2$ | 162583 | mmm | 4 | 16 | 0 | 0 | 17 | 9.1 | 13.6 | 0 |
| $Li_2BPt_3$ | 156466 | 432 | 4 | 24 | 0 | 0 | 7 | 3.9 | 5.2 | 0 |
| $LiBiO_3$ | 82277 | mmm | 8 | 40 | 1.2 | 0 | 3032 | 16.1 | 5.5 | 0 |
| $Sr_2P_7Br$ | 429306 | 23 | 4 | 40 | 2.8 | 0 | 98 | 61.6 | 26.4 | 0 |
| CaPdSi | 69790 | 2/m | 4 | 12 | 0 | 2 | 90 | 9.0 | 19.8 | 0 |
| $CdI_2O_6$ | 1397 | 222 | 4 | 36 | 4.5 | 0 | 3030 | 9.0 | 19.8 | 0 |
| $Cs_2SbCl_6$ | 49706 | 4/mmm | 4 | 36 | 1.6 | 0 | 4 | 21.7 | 12.2 | 0 |
| $LiWCl_6$ | 409938 | 3 | 4 | 32 | 0 | 12 | 739 | 8.1 | 14.4 | 0 |
| $TlGaS_2$ | 157537 | 2/m | 8 | 32 | 2.4 | 0 | 3 | 0.0 | 7.1 | 0 |
| $TlGaSe_2$ | 1573 | m | 8 | 32 | 2.1 | 4 | 737 | 32.6 | 28.1 | 0 |
| HfNbP | 75009 | mmm | 4 | 12 | 0 | 0 | 32 | 3.7 | 2.6 | 0 |
| $Zr_2Pd_2In$ | 107332 | 4/mmm | 4 | 20 | 0 | 135 | 4 | 3.7 | 6.9 | 0 |
| IrSbTe | 640967 | 23 | 4 | 12 | 1.3 | 0 | 22 | 9.9 | 6.0 | 0 |
| $Li_2TeSe_3$ | 415121 | 2/m | 4 | 24 | 0.19 | 0 | 161 | 12.7 | 10.1 | 0 |
| $PbN_2O_6$ | 174004 | $m\bar{3}$ | 4 | 36 | 5.5 | 0 | 8 | 0.2 | 8.6 | 0 |
| $Na_3SbTe_3$ | 75513 | 23 | 4 | 28 | 2.0 | 16 | 190 | 14.1 | 12.3 | 0 |
| $TlSbO_3$ | 4123 | $\bar{3}m$ | 4 | 20 | 3.3 | 135 | 27 | 5.2 | 7.5 | 0 |
| $Tl_3PS_4$ | 201062 | mmm | 4 | 32 | 2.7 | 0 | 21 | 10.2 | 12.4 | 0 |
| $BaIn_2Te_4$ | 41168 | mmm | 2 | 14 | 1.6 | 0 | 2802 | 10.1 | 16.9 | 0 |
| $Rb_2ZrTe_3$ | 410735 | 2/m | 4 | 24 | 0.14 | 0 | 25 | 44.4 | 8.6 | 0 |
| $Ag_2HgO_2$ | 280333 | 422 | 4 | 20 | 1.3 | 0 | 321 | 22.1 | 16.3 | 0 |
| RbAgO | 40155 | 4/mmm | 4 | 12 | 2.7 | 0 | 3577 | 19.8 | 9.7 | 0 |
| $AsNb_3Te_3$ | 79934 | 6/m | 2 | 14 | 0 | 0 | 408 | 48.2 | 35.5 | 0 |
| $PbSnS_3$ | 23462 | mmm | 4 | 20 | 1.6 | 7 | 1528 | 35.1 | 16.0 | 0 |
| $CaPS_3$ | 405192 | 2/m | 4 | 20 | 4.1 | 0 | 131 | 4.9 | 11.9 | 0 |
| $HfSiO_4$ | 31177 | 4/mmm | 2 | 12 | 7.0 | 0 | 35 | 14.9 | 27.2 | 0 |
| $NaScS_2$ | 644971 | $\bar{3}m$ | 1 | 4 | 2.7 | 0 | 68 | 5.0 | 13.7 | 0 |
| $MgTa_2O_6$ | 202688 | 4/mmm | 2 | 18 | 4.2 | 0 | 9 | 2.4 | 7.3 | 0 |
| $Mg_2SiO_4$ | 15627 | mmm | 4 | 28 | 6.4 | 0 | 152 | 1.8 | 3.0 | 0 |
| $Na_2SO_3$ | 31816 | $\bar{3}$ | 2 | 12 | 6.4 | 41 | 761 | 3.0 | 7.0 | 0 |
| $Ta_4SiTe_4$ | 40207 | mmm | 4 | 36 | 0 | 0 | 860 | 11.6 | 9.6 | 0 |
| $KBS_2$ | 79614 | $\bar{3}m$ | 6 | 24 | 3.8 | 0 | 113 | 10.1 | 17.3 | 0 |
| AlCaSi | 155193 | 6 | 6 | 18 | 0 | 0 | 5 | 15.3 | 47.9 | 1.1 |
| $NaPt_2Se_3$ | 78788 | 6mm | 4 | 24 | 1.6 | 0 | 989 | 2.2 | 7.6 | 1.2 |
| $BaAl_2Si_2$ | 249559 | mmm | 4 | 20 | 0 | 0 | 4 | 40.6 | 20.7 | 4.6 |
| $BaGe_2S_5$ | 66868 | $m\bar{3}m$ | 4/16 | 32/ 128 | 3.2 | 0 | 131 | 43.3 | 12.7 | 5.3/−2.3* |
| $RbPSe_3$ | 173419 | 32 | 6 | 30 | 2.0 | 0 | 8 | 57.6 | 15.3 | 5.7 |
| $CsIn_3O_5$ | 23630 | mmm | 4 | 36 | 2.8 | 0 | 2017 | 1.2 | 6.0 | 6.8 |
| $KAsSe_2$ | 65297 | 1 | 4 | 16 | 2.3 | 0 | 4909 | 40.9 | 34.2 | 8.0 |
| CaAsPt | 60828 | 4mm | 6 | 18 | 0 | 0 | 10 | 7.4 | 4.0 | 9.0 |
| $AlBiCl_6$ | 414261 | 2/m | 4 | 32 | 5.0 | 0 | 23 | 14.3 | 7.6 | 13.0 |
| $Na_3SbO_3$ | 23346 | $\bar{4}3m$ | 4/8 | 28/ 56 | 4.1 | 0 | 4958 | 2.4 | 8.1 | 13.3/0.0* |
| $Sb_2OS_2$ | 12120 | $\bar{1}$ | 8 | 40 | 1.8 | 1 | 172 | 38.0 | 16.1 | 18.6 |
| $Na_2AuSn_3$ | 107556 | 6/mmm | 4 | 24 | 0 | 0 | 2357 | 36.6 | 8.3 | 25.3 |
| $SnGeS_3$ | 411241 | 2/m | 4 | 20 | 1.9 | 0 | 4993 | 75.5 | 227.3 | 30.0 |
| YPdGe | 391466 | mm2 | 6/12 | 18/ 36 | 0 | 0 | 680 | 21.6 | 9.9 | 32.9/0.0* |
| $Sr_2Pt_3In_4$ | 410703 | $\bar{6}$m2 | 4 | 36 | 0 | 0 | 3575 | 83.5 | 140.0 | 36.6 |